\documentclass[prd,preprint,nofootinbib]{revtex4}
\renewcommand{\=}[1]{\bar{#1}}
\newcommand{\sect}[1]{\section{#1}\setcounter{equation}{0}}

\newcommand{\OL}[1]{ \hspace{1pt}\overline{\hspace{-1pt}#1
   \hspace{-1pt}}\hspace{1pt} }
\newcommand{\be}{\begin{equation}}
\newcommand{\ee}{\end{equation}}
\newcommand{\bea}{\begin{eqnarray}}
\newcommand{\eea}{\end{eqnarray}}
\newcommand{\beas}{\begin{eqnarray*}}
\newcommand{\eeas}{\end{eqnarray*}}
\newcommand{\br}{\begin{array}{cc}}
\newcommand{\bc}{\begin{array}{c}}
\newcommand{\ea}{\end{array}}

\renewcommand{\b}[1]{\bar{#1}}

\newcommand{\del}{\partial}

\newcommand{\bi}{\bar{\imath}}
\newcommand{\bj}{\bar{\jmath}}
\renewcommand{\ap}{\alpha^\prime}

\newcommand{\N}{{\cal N}}
\renewcommand{\mathsf}[1]{\cal #1}
\newcommand{\skyp}[1]{}

\begin{document}
\preprint{NSF-ITP-01-77}

\title{${\cal N}=3$ Warped Compactifications}

\author{Andrew R. Frey}
\email{frey@vulcan.physics.ucsb.edu}
\affiliation{Department of Physics\\ University of California\\ 
Santa Barbara, CA 93106}

\author{Joseph Polchinski}
\email{joep@itp.ucsb.edu}
\affiliation{Institute for Theoretical Physics\\ University of California\\ 
Santa Barbara, CA 93106-4030}

\begin{abstract}
Orientifolds with three-form flux provide some of the simplest string
examples of warped compactification.  In this paper we show that some
models of this type have the unusual feature of $D=4$, $\N=3$ spacetime
supersymmetry.  We discuss their construction and low energy physics.
Although the local form of the moduli space is fully determined by
supersymmetry, to find its global form requires a careful study of the BPS
spectrum.
\end{abstract}

\date{\today}
\maketitle

\section{Introduction}\label{s:intro}

Warped compactifications are of great interest, due to the
observation of Randall and Sundrum that warping in a higher
dimensional space can produce a hierarchy of four-dimensional
scales~\cite{Randall:1999ee,Randall:1999vf}.
Becker and Becker~\cite{Becker:1996gj} described a large class of
warped three-dimensional M theory compactifications, in which four-form
flux is the source for the warp factor.  By duality these give rise to
warped four-dimensional IIB compactifications, with three-form fluxes as
the source~\cite{Dasgupta:1999ss,Gukov:1999ya}.

In this paper we study some particularly simple examples of this
type, which as we will show have $D=4$, ${\cal N}=3$ supersymmetry.
These are of interest in part because of the rarity of ${\cal N}=3$
supersymmetry, but also because the supersymmetry strongly constrains
their moduli spaces.  The small-radius
behavior of warped compactifications is likely to be quite complicated for
${\cal N} \leq 2$, as the warping becomes large in this limit and the
application of
$T$-duality (or mirror symmetry) is complicated by the warping and fluxes.
Also, such compactifications are intrinsically nonperturbative, in that the
dilaton is fixed at a nonzero value. However, with ${\cal N}=3$
supersymmetry the local form of the moduli space is completely determined,
and we can hope to deduce the global structure.

In \S \ref{s:back} we describe these solutions, all of which are based on the
$T^6/{\bf Z}_2$ orientifold, and discuss their supersymmetry.  An
interesting subtlety arises with the flux quantization.
In \S3 we study various aspects of the low energy physics --- the massless
spectrum, and the metric on moduli space --- and show that it is
consistent with the constraints of ${\cal N} = 3$ supersymmetry.
We argue that the breaking of $\N=4$ to $\N=3$ should appear to be
spontaneous in the large radius limit.  In \S4
we consider the duality groups.  Because of the $H_{(3)}$ flux and the
finite $g_{\rm s}$, we have no tools to determine these directly, and so
must try to deduce their form based on the spectrum of BPS states.  We
find that, even though it may be possible to view the duality group as a
spontaneous breaking of the $\N=4$ dualities, the symmetry beaking is 
not straightforward.


While this work
was in progress we learned of related work on $T^6/Z_2$
orientifolds with flux~\cite{Kachru:2002he}.  We are grateful to those
authors for communications.

\sect{${\cal N} = 3$ orientifolds}\label{s:back}

In this section we describe the specific orientifold solutions with
three-form flux, and determine their supersymmetry.  This overlaps the
discussion in ref.~\cite{Dasgupta:1999ss}; the
$T^6/{\bf Z}_2$ orientifold was discussed briefly there, in its M theory
avatar $T^8/{\bf Z}_2$.

In \S \ref{s:oproject} we determine the action of the $T^6/{\bf
Z}_2$ orientifold projection on the fields.  In \S \ref{s:fluxquant}
we discuss the quantization of three-form flux, which has an interesting
subtlety. In \S \ref{s:bianchi} we describe the solution to the
Bianchi identities and equations of motion.  In \S \ref{s:examples}
we identify a particularly simple class of models, in which only one
complex component of the flux is nonvanishing.  In \S \ref{s:susy3}
we study the supersymmetry of these models and show that there are ${\cal
N}=3$ unbroken supersymmetries.

\subsection{Orientifold projection}\label{s:oproject}

All examples that we consider are based on the $T^6/{\bf Z}_2$ orientifold.
Greek indices denote the noncompact directions $0,\ldots,3$, lower
case roman indices denote the compact directions $4, \ldots,9$, and capital
roman indices denote all directions $0,\ldots,9$.
The coordinates $x^m$ are each taken to be periodic
with period $2\pi$, and the ${\bf Z}_2$ is simultaneous reflection of all
compact coordinates $x^m$,
\begin{equation}
R: \quad (x^4, x^5, x^6, x^7, x^8, x^9) \to  (-x^4, -x^5, -x^6, -x^7, -x^8,
-x^9)\ .
\end{equation}
For now we take the toroidal metric to be rectangular,
\begin{equation}
\widetilde{ds}^2 = \sum_{m=4}^9 r_m^2 dx^m dx^m\ ;
\label{rect}
\end{equation}
we will relax this in \S 3.

The action of the orientifold
${\bf Z}_2$ can be derived by using $T$-duality to the type I theory, where
$g_{MN}$, $C_{(2)}$, and
$\Phi$ are even under world-sheet parity $\Omega$ and $B_{(2)}$,
$C \equiv C_{(0)}$, and $C_{(4)}$ are odd.
Alternately, one may derive it by noting that the orientifold ${\bf Z}_2$
must include a factor of $(-1)^{{\bf F}_L}$, where ${\bf F}_L$ is the
spacetime fermion number carried by the left-movers:
${\cal R} \equiv \Omega R (-1)^{{\bf
F}_L}$~\cite{Dabholkar:1996pc,Sen:1996vd}.  This is necessary in order that
it square to unity,
\begin{equation}
{\cal R}^2 =
\Omega^2 R^2 (-1)^{{\bf F}_L + {\bf F}_R} =1\ .
\end{equation}
Note that $\Omega^2 = 1$, as $\Omega$ acts as $\pm 1$ on all fields.
$R$ is equivalent to a rotation by $\pi$ in each of three planes, so $R^2$
is a rotation by $2\pi$ in an odd number of planes and therefore equal to
$(-1)^{\bf F}$.

By either means one finds that ${\bf Z}_2$ acts on the various fields as
follows:
\begin{eqnarray}
\mbox{even:}&&
g_{\mu\nu}\ ,\ g_{mn}\ ,\ B_{\mu m}\ ,\ C_{\mu m}\ , \
C_{mnpq}\ ,\ C_{\mu\nu mn}\ ,\ C_{\mu \nu\lambda\rho}\ ,\
\Phi\ ,\  C \ ;\nonumber\\
\mbox{odd:}&& g_{\mu m}\ ,\ B_{\mu \nu}\ ,\ B_{mn}\ ,\ C_{\mu \nu} \
,\ C_{mn}\ ,\  C_{\mu mnp}\ ,\ C_{\mu\nu\lambda m}\ .
\label{reflect}
\end{eqnarray}
It follows that the fluxes $H_{mnp}$ and $F_{mnp}$ are even, and so
constant three-form fluxes are allowed.

\subsection{Flux quantization}\label{s:fluxquant}

The three-form fluxes must be appropriately quantized.  The usual
quantization conditions are\footnote{We follow the conventions of
ref.~\cite{Polchinski:2000uf}.}
\begin{equation}
\frac{1}{2\pi \alpha'} \int_{C} H_{(3)} \in 2\pi {\bf Z}\ ,\quad
\frac{1}{2\pi \alpha'} \int_{C} F_{(3)} \in 2\pi {\bf Z}
\label{gquant}
\end{equation}
for every three-cycle $C$.  However, the
orientifold presents some subtleties.

Consider first $T^6$ compactification.  Letting $C$
run over all $T^3$'s, one finds that constant fluxes
\begin{equation}
H_{mnp} = \frac{\alpha'}{2\pi} h_{mnp}\ ,\quad F_{mnp} =
\frac{\alpha'}{2\pi} f_{mnp}\ ;\quad h_{mnp}\ ,\ f_{mnp} \in {\bf Z}
\label{3quant}
\end{equation}
are allowed.  Any cycle on the covering space $T^6$ descends to a cycle on
$T^6/{\bf Z}_2$, so the
conditions~(\ref{3quant}) are still necessary.  In addition, there are new
3-cycles on the coset space, such as
\begin{equation}
0 \leq x^4 \leq 2\pi\ ,\quad
0 \leq x^5 \leq 2\pi\ ,\quad
0 \leq x^6 \leq \pi\ ,\quad
x^7 = x^8 = x^9 = 0\ . \label{halfc}
\end{equation}
The conditions~(\ref{gquant}) on this cycle\footnote
{The cycle~(\ref{halfc}) is unoriented, but the three-form fluxes can be
integrated on it because they have odd intrinsic parity.}
would appear to
require that
$h_{456}$ and $f_{456}$ be even, and similarly for all other components.
However, we claim that $h_{mnp}$ and $f_{mnp}$ can still be arbitrary odd
or even integers.

To understand this, consider first the reduced problem of a charge moving
in a constant magnetic field $F_{56}=F$ on a torus $0 \leq x^{5,6} \leq
2\pi$. Let us work in the gauge
\begin{equation}
A_5 = 0\ ,\quad A_6 = F x^5\ .
\end{equation}
The gauge field is periodic up to a gauge transformation,
\begin{equation}
A_m (x^5 + 2\pi, x^6) = A_m (x^5 , x^6) + \partial_m \lambda_5\ ,
\quad
A_m (x^5, x^6 + 2\pi) = A_m (x^5 , x^6) + \partial_m \lambda_6\ ,
\end{equation}
with $\lambda_5 = 2\pi F x^6$ and $\lambda_6 = 0$.  Similarly a field of unit
charge satisfies
\begin{equation}
\psi (x^5 + 2\pi, x^6) = e^{i \lambda_5} \psi (x^5 , x^6) \ ,
\quad
\psi (x^5, x^6 + 2\pi) = e^{i \lambda_6} \psi (x^5 , x^6)\ .
\label{gtran}
\end{equation}
The consistency of defining $\psi (x^5+2\pi, x^6 + 2\pi)$ implies the Dirac
quantization
\begin{equation}
F = f / 2\pi\ ,\quad f \in {\bf Z}\ .
\end{equation}
In other words,
\begin{equation}
\int_{T^2} F_{56} = 2\pi f \in 2\pi {\bf Z}\ .
\end{equation}
Now let us form the orbifold $T^2 / {\bf Z}_2 = S^2$ by identifying
$(x^5,x^6)$ with $(-x^5,-x^6)$.  For any value of
$f$ we can define the quantum mechanics for the charged particle on the
coset space simply by restricting to wavefunctions such that\footnote
{We have chosen a gauge in which $A_m$ is explicitly ${\bf Z}_2$ symmetric,
so no gauge transformation is needed.}
\begin{equation}
\psi (-x^5, -x^6) = + \psi (x^5 , x^6) \ . \label{gref}
\end{equation}
However, the integral of $F_{56}$ over
$S^2$ is half of the integral over $T^2$, so for $f$ odd the flux
is not quantized.

To see how this can make sense, note that there are four fixed points
$(x^5,x^6) = (0,0),\ (\pi,0),\ (0,\pi),\
(\pi,\pi)$.  At the first three, the
periodicities~(\ref{gtran}) and~(\ref{gref}) are compatible, but at
$(\pi,\pi)$ they are incompatible and the wavefunction must
vanish.  If we circle this fixed point, from
$(\pi-\epsilon,\pi)$ to the identified point
$(\pi+\epsilon,\pi)$, the wavefunction is required to change sign:
there is a half-unit of magnetic flux at the fixed point
$(\pi,\pi)$.  Thus the Dirac quantization condition is in
fact satisfied.

Of course, the fixed point $(\pi,\pi)$ is not special: the quantization
condition is satisfied if there is a half-unit of flux at any one fixed
point, or at any three.  Similarly for $f$ even there can be
half-integer flux at zero, two, or four fixed points.  In each case
there are eight configurations, which can be obtained in the orbifold
construction by including discrete Wilson lines on the torus, and a
discrete gauge transformation in the orientifold projection.

This analysis extends directly to the quantum mechanics of an F-string or
D-string wrapped in the 4-direction, moving in the fluxes $H_{456}$ and
$F_{456}$.  This is consistent for any integers $h_{456}$ and $f_{456}$,
but if either of these is odd then there must NS-NS or R-R flux at
some fixed points, for example all those with $x^4 = x^5 =
x^6 =\pi$.
Indeed, there are four kinds of O3 plane, distinguished by the presence or
absence of discrete NS-NS and R-R fluxes~\cite{Witten:1998xy}; for
recent reviews see ref.~\cite{Hanany:2000fq,Bergman:2001rp}.  The
cycle~(\ref{halfc}), and each of the others obtained from it by a rotation
of the torus, contains four fixed points.  If the NS-NS flux through the
cycle is even (odd) then an even (odd) number of the fixed points must have
discrete NS-NS flux, and correspondingly for the R-R flux.

\subsection{Bianchi identities and field equations}\label{s:bianchi}

The Bianchi identities for the three-form flux, $dH_{(3)} = dF_{(3)} = 0$,
are trivially satisfied by constant fluxes. The Bianchi identity for the
five-form flux is
\begin{equation}
d \tilde F_{(5)} = (2\pi)^4 \alpha'^2 \rho^{\rm loc}_3 dV_\perp + H_{(3)}
\wedge F_{(3)}\ ,
\label{bian5}
\end{equation}
where $\omega^{\rm loc}_3$ is the D3-brane density from localized
sources and $dV_\perp$ is the transverse volume form.
The localized sources that we will consider are D3-branes and the
various types of O3-plane.  An O3-plane without discrete flux
has D3 charge $-\frac{1}{4}$~\cite{Polchinski:1995mt}, while an O3-plane
with either discrete flux, or with both, has D3 charge
$+\frac{1}{4}$~\cite{Witten:1998xy,Hanany:2000fq,Bergman:2001rp}.
The integrated Bianchi
identity then gives the tadpole cancellation condition
\begin{equation}
N + \frac{1}{2} \tilde N + \frac{1}{2 \cdot 3! \cdot 3!}
\hat\epsilon^{mnpqrs} h_{mnp} f_{qrs} = 16
\ . \label{biint}
\end{equation}
Here $N$ is the total number of D3-branes, $\tilde N$ is the
total number of O3 planes with any discrete flux, and
$\hat\epsilon^{456789} = 1$.  The factor of $\frac12$ in the flux term
arises because the orientifold has half the volume of the original torus.

We are interested in compactifications to
four-dimensional Minkowski space with supergravity fields plus D3-branes
and O3-planes.  In ref.~\cite{Giddings:2001yu} it is shown that all such
solutions must be of `smeared D3' form~\cite{Grana:2000jj,Gubser:2000vg},
which is dual to the M theory ansatz of ref.~\cite{Becker:1996gj}.
That is, the flux
\begin{equation}
G_{(3)} = F_{(3)} - \tau H_{(3)}\ , \quad \tau = C + i e^{-\Phi}\ ,
\end{equation}
must be imaginary self-dual,
\begin{equation}
\frac{1}{3!} \epsilon_{mnp}{}^{qrs} G_{qrs} = i G_{mnp}\ . \label{imsd}
\end{equation}
This flux
behaves as an effective D3-brane source for the remaining fields, which
are therefore of black 3-brane form~\cite{Horowitz:1991cd}
\begin{eqnarray}
\tau &=& \mbox{constant} \equiv C + \frac{i}{g_{\rm s}}\ ,\nonumber\\
ds_{\rm string}^2 &=& Z^{-1/2} \eta_{\mu\nu} dx^\mu dx^\nu + Z^{1/2}
\tilde g_{mn} dx^m dx^n\ , \nonumber\\
\tilde F_{(5)} &=&  (1 +
{*})d \chi_{(4)}\ ,\quad \chi_{(4)} = \frac{1}{g_{\rm s} Z} dx^0 \wedge dx^1
\wedge dx^2 \wedge dx^3
\ .
\end{eqnarray}
The warp factor $Z$ is determined by
\begin{equation}
\label{warping}
-\widetilde{\nabla}^2  Z = (2\pi)^4 \alpha^{\prime 2} g_{\rm s}
\tilde\rho_{3} + \frac{g_{\rm s}}{12} G_{mnp}
\OL G^{\widetilde{mnp}}\ ,
\end{equation}
where a tilde denotes the use of the unwarped metric~(\ref{rect}).
This is consistent provided that the net D3 charge~(\ref{biint})
vanishes, and the Bianchi identity~(\ref{bian5}) and the field equations
are then satisfied.

As discussed in \cite{Greene:2000gh}, the warp factor can be obtained from
(\ref{warping}) by the method of images.  Under rescaling of the unwarped
transverse metric, $\tilde g_{mn} \to
\lambda^2 \tilde g_{mn}$, the right-hand side of eq.~(\ref{warping}) scales
as $\lambda^{-6}$ (there is a factor of $\tilde g^{-1/2}$ in
$\rho_3$), while $\tilde\nabla^2$ scales as
$\lambda^{-2}$.  It follows that in the large radius
limit $Z = 1 + O(\lambda^{-4})$ and the warping becomes negligible.  On the
other hand, at small radius the warping is significant.  Thus we might
expect that in general the small radius region of moduli space is
significantly modified --- for example, the AdS radius of the warped region
remains finite even as the radius of the unwarped manifold is taken to
zero.  Note also that due to the negative charge of the orientifold planes,
the warp factor becomes negative and unphysical near the
${\bf Z}_2$ fixed points.  Since the region of unphysical behavior is
smaller than the string scale, the geometry cannot be taken literally, but
it again suggests that the small-radius limit may be
complicated.\footnote {This remark is due to S. Sethi.}  However, for
the highly supersymmetric cases that we consider the small-radius limit
is highly constrained.

\subsection{Examples}\label{s:examples}

There are many solutions based on the $T^6/{\bf Z}_2$ orientifold,
distinguished by the three-form flux quanta and the discrete fluxes at
orientifold points.  Even with vanishing three-form fluxes there are many
solutions to the tadpole cancellation condition~(\ref{biint}) and the
three-form flux quantization conditions.  One extreme is to have 16
D3-branes and no discrete flux~\cite{Verlinde:1999fy}, which is the familiar
$T$-dual to the type I theory on $T^6$.  The other extreme is to have no
D3-branes and 32 fixed points with discrete flux.  For example, the
configuration with discrete R-R flux at all fixed points in the plane $x^4 =
0$ satisfies the quantization conditions and is $T$-dual to a type I
compactification without vector structure~\cite{Witten:1998bs}.
In these cases the supersymmetry is
$D=4$, ${\cal N} = 4$.

For simplicity we will restrict attention to a
limited set of three-form flux configurations, where the nonzero fluxes are
\begin{eqnarray}
h_{456} = -h_{489} = -h_{759} = -h_{786} &\equiv& h_1\ ,\nonumber\\
f_{456} = -f_{489} = -f_{759} = -f_{786} &\equiv& f_1\ ,\nonumber\\
h_{789} = -h_{756} = -h_{486} = -h_{459} &\equiv& h_2\ ,\nonumber\\
f_{789} = -f_{756} = -f_{486} = -f_{459} &\equiv& f_2\ , \label{fluxes}
\end{eqnarray}
and $f_{1,2}$ and $h_{1,2}$ are integers. The duality condition~(\ref{imsd})
implies that the
$T^6$ is the product of three square $T^2$'s,
\begin{equation}
r_4 = r_7 \ ,\quad
r_5 = r_8 \ ,\quad
r_6 = r_9 \ ,
\label{pairrad}
\end{equation}
and that the string coupling is
fixed in terms of the integer fluxes,
\begin{equation}
\tau = \frac{f_2 - i f_1}{h_2 - i h_1}\ . \label{taufix}
\end{equation}
This is therefore an intrinsically nonperturbative
solution of IIB string theory.  It can be studied at large
radius using supergravity, which becomes classical at low energy, but to
understand the physics at small radius a high degree of supersymmetry will
be essential.  The tadpole cancellation condition is
\begin{equation}
N + \frac{1}{2} \tilde N = 16 - 2(h_1 f_2 - h_2 f_1)
\leq 16.
\end{equation}
the last inequality follows from the duality condition (\ref{taufix}).

This configuration of fluxes has the simple feature that in terms of the
complex coordinates
\begin{equation}\label{complex}
w^1 = \frac{x^4 + i x^7}{\sqrt 2}\ ,\quad
w^2 = \frac{x^5 + i x^8}{\sqrt 2}\ ,\quad
w^3 = \frac{x^6 + i x^9}{\sqrt 2}\ ,
\end{equation}
there is a single component
\begin{equation}
G_{\bar 1 \bar 2 \bar 3} = \frac{\sqrt 2\ap}{\pi} (f_1 - \tau h_1)\ .
\label{03}
\end{equation}
That is, $G_{mnp}$ is a $(0,3)$-form.
Such solutions will be the focus of the remainder of this paper.
The unwarped metric is
\begin{equation}
\widetilde{ds}^2 = 2\sum_{i=1}^3 r_{i+3}^2 dw^i d\bar w^{\bar\imath} \
,\quad
\tilde{g}_{i\bj} = r_{i+3}^2 \delta_{i\bj}\ .
\end{equation}

If we restrict to even $f_{1,2}$ and $h_{1,2}$, and to O3-planes without
flux, then it is easy to list all solutions, up to rotations and dualities:
\begin{equation}
\begin{array}{rll}
\mbox{(A)}\ & h_1 = f_2 = 2\ ,\ h_2 = f_1 = 0\ :\quad&
N = 8\ ,\ g_{\rm s} =1\ ,\ C=0\ ;
\nonumber\\[3pt]
\mbox{(B)}\ &h_1 = 2\ ,\ f_2 = 4\ ,\ h_2 = f_1 = 0\ :\quad&
N = 0\ ,\ g_{\rm s} =\mbox{$\frac{1}{2}$}\ ,\ C=0\ ;
\nonumber\\[3pt]
\mbox{(C)}\ &h_1 = -h_2 = f_1 = f_2 = 2\ :\quad&
N = 0\ ,\ g_{\rm s} =1\ ,\ C=0\ .
\end{array}
\end{equation}
For example, the solution $h_1 = f_2 = 2$, $h_2 = 0$, $f_1 = 2m$,
with $N = 8$ and $\tau = i + m$, is $S$-dual to case (A).

With odd fluxes and discrete flux on the O3-planes the number of solutions
is large.  One example is $h_1 = 1$, $f_2 = 4$, $h_2 = f_1 = 0$, $N=0$,
$g_{\rm s} = \frac{1}{4}$, $C=0$, with
discrete NS-NS flux at the 16 fixed points at which exactly one of the
following four conditions holds: [$x^4=x^5=x^6=0$],\ \
[$x^4=x^8=x^9=0$],\ \ [$x^7=x^5=x^9=0$],\ \ [$x^7=x^8=x^6=0$].

In the notation of ref.~\cite{Dasgupta:1999ss} (eq.~(3.18) and (3.19) of
version 3), the ansatz~(\ref{fluxes},\ \ref{03}) corresponds to solutions
with only
$A$ nonvanishing; in particular case (C) is the solution $A = 1+i$.  The
condition (3.18) in ref.~\cite{Dasgupta:1999ss} is equivalent to $f_{mnp}$
and $h_{mnp}$ being even in our notation.

\subsection{Supersymmetry counting}\label{s:susy3}

The supersymmetry of this class of IIB solutions was discussed in
refs.~\cite{Grana:2000jj,Gubser:2000vg}.  Aside from the three-form
fluxes, the background is a distribution of black 3-branes.  Therefore the
supersymmetries of the black/D3-brane,
\begin{equation}
SO(3,1) \times SO(6):\quad \varepsilon = \zeta \otimes \chi\ ,\quad
\Gamma_{(4)}
\zeta = + \zeta\ ,\quad
\Gamma_{(6)} \chi = - \chi\ , \label{spinor}
\end{equation}
are broken only by terms that are linear in the three-form fluxes.
Using the supersymmetry transformations from
refs.~\cite{Schwarz:1983wa,Schwarz:1983qr,Howe:1984sr}, the unbroken
supersymmetries are those that satisfy
\begin{equation}
G \chi = G \chi^* = G \gamma^m \chi^* = 0\ ,\quad G \equiv \frac{1}{6}
G_{mnp} \gamma^{mnp} = G_{\bar 1 \bar 2 \bar 3} \gamma^{\bar 1 \bar 2 \bar 3}
\ .  \label{vary}
\end{equation}
A spinor $\chi$ of chirality~(\ref{spinor}) is either
$\chi_0$, where
\begin{equation}
\gamma^{\bar\imath} \chi_0 = 0 \quad \mbox{(all $i$)}\ ,
\end{equation}
or one of the three spinors $\gamma^{ij} \chi_0$.  One readily verifies
that for the latter three spinors the conditions~(\ref{vary}) are satisfied
and so the unbroken supersymmetry is $D=4$, ${\cal N}=3$.  The number ${\cal
N}$ of solutions to the conditions~(\ref{vary}) can be any of 0, 1, 2, 3,
and 4 (the last is for vanishing fluxes); all but the case ${\cal N} = 3$
have been discussed in the previous work.

The ${\cal N}=3$ supersymmetry can be understood simply as follows.  The
condition for an unbroken supersymmetry is that the flux $G_{(3)}$ be of type
$(2,1)$ and primitive~\cite{Grana:2000jj,Gubser:2000vg}.  The orientifold
has several complex structures. If we choose the coordinates
\begin{equation}
(z^1,z^2,z^3) = (w^1,\bar w^{\bar 2},\bar w^{\bar 3}) \label{z1}
\end{equation}
then the nonzero flux $G_{\bar z^{\b{1}} z^2 z^3}$ is indeed $(2,1)$ and
primitive.  There are obviously two other such choices,
\begin{equation}
(z^1,z^2,z^3)' = (\bar w^{\bar 1},w^2,\bar w^{\bar 3})\ ,\quad
(z^1,z^2,z^3)'' = (\bar w^{\bar 1},\bar w^{\bar 2},w^{3})\ . \label{z2}
\end{equation}
Each of these three complex structures leads to an unbroken supersymmetry.

${\cal N}=3$ supersymmetry is unfamiliar but not unknown.  Previous
examples have been constructed as asymmetric orbifolds in type II
theory~\cite{Ferrara:1989nm}, breaking half of the supersymmetry on one
side and three-fourths on the other.  The
${\cal N}=3$ matter multiplet (helicities $1, \frac{1}{2}^3, 0^3,
-\frac{1}{2}$) plus its $CPT$ conjugate form an ${\cal N}=4$ matter
multiplet, but the supergravity multiplet (helicities $2, \frac{3}{2}^3,
1^3, \frac{1}{2}$ plus $CPT$ conjugates) is distinct.  In the
global case the renormalizable interactions are the same as for ${\cal
N}=4$, but there are presumably higher-dimension operators allowed by
${\cal N}=3$ but not ${\cal N}=4$.  The ${\cal N}=3$ supergravity was
constructed in ref.~\cite{Castellani:1986ka}.  Like ${\cal N} = 4$, the
moduli space is a coset and its local form is completely determined,
\begin{equation}
\frac{U(3,n)}{U(3) \times U(n)}\ , \label{metmod}
\end{equation}
where $n$ is the number of matter multiplets.  Including the vectors in the
supergravity multiplet, the gauge symmetry is $U(1)^{n+3}$.

\sect{Low energy effective theory}

In this section we analyze the massless spectra of the models described in
the previous section, to verify the structure required by ${\cal N} = 3$
supergravity: with the supergravity multiplet plus $n$ matter multiplets,
there must be $6n$ moduli and $n+3$ vectors.
We also verify, in the large-radius limit, that the metric
on moduli space has the expected form~(\ref{metmod}).
Note that, because $g_{\rm s}$ is fixed to be of order one, we cannot use
string perturbation theory to study these models.  The one tool we have is
low energy supergravity, which is valid in the large-radius limit.  In
this ${\cal N}= 3$ case there is enough supersymmetry to extrapolate to
the full moduli space, but for ${\cal N}\leq 2$ it will be very
difficult to analyze the full moduli space.

\subsection{Moduli}

The massless scalars arise from the zero modes of the ${\bf Z}_2$-even
scalars in (\ref{reflect}), namely $g_{mn}$, $C_{mnpq}$, $\Phi$ and $C$.
However, not all of these are moduli, as the
fluxes lift some of the directions of moduli
space~\cite{Dasgupta:1999ss,Gukov:1999ya,Greene:2000gh}.  For example,
we have already seen that the dilaton and R-R scalar are fixed. Their
potential arises from the three-form flux and the resulting mass-squared is
of order
\begin{equation}
G_{mnp} G^{mnp} \sim \frac{\alpha'^2}{R^6}\ .\label{fluxmass}
\end{equation}
We have assumed that all radii of the torus are of order $R$, so that
$g_{mn} \sim R^2$, and have used the quantization
conditions~(\ref{gquant}).

Now consider the scalars $g_{mn}$.  These are partly fixed by the
self-duality condition~(\ref{imsd}), through the dependence of the
$\epsilon$-tensor on $g_{mn}$.  The zero mode of the three-form flux is
fixed by the quantization conditions, so $G_{mnp}$ remains a
$(0,3)$-form in the $w$ coordinates.  The metric $g_{mn}$ must therefore be
Hermitean in these coordinates, else there will be nonzero components
$\epsilon_{\bar\imath\bar\jmath k}{}^{\bar\imath'\bar\jmath' \bar k'}$.
The self-duality condition is satisfied for any Hermitean metric
$g_{i\bar\jmath}$.  Thus, in terms of the $w$ coordinates, the complex
structure moduli are frozen while the K\"ahler moduli remain free.  In
terms of any of the supersymmetric complex structures~(\ref{z1}, \ref{z2})
these are a mix of K\"ahler and complex structure moduli.

The remaining bulk scalars are those from the four-form
potential
$C_{mnpq}$.  The periodicity conditions on this potential are slightly
involved, and so the analysis is set aside below.
The conclusion is that there is a field $\tilde c_{mnpq}$ which is periodic
and which appears in the field strength only through its exterior
derivative.  A constant shift of this field is then a new solution to the
equations of motion.  However, some of these are gauge-equivalent to the
unshifted solution.  It is shown below that the gauge variation
around a given background is
\begin{equation}
\delta \tilde c_{(4)} = d\tilde\chi +i
(\OL{\lambda}_A \wedge \hat G_{(3)} - \lambda_A \wedge
\hat{\OL{G}}_{(3)})/2\,{\rm Im}(\tau)\ , \label{c4varx}
\end{equation}
with $\tilde \chi$ periodic and $\lambda_A$ a complex
one-form.  Since the background $\hat G_{(3)}$ is a
$(0,3)$ form, the $(1,3)$ and $(3,1)$ parts of $\tilde c_{(4)}$ can be
gauged away.  The $(2,2)$ parts $\tilde c_{ij\bar k\bar l}$ are the
moduli.

Finally, there is no restriction on the positions of any D3-branes that
might be present, so their world-volume scalars are also
moduli.  It will be convenient to write these in complex form,
as $W_I^i$, $\OL W_I^{\bj}$ where $I$ labels the D3-brane (perturbatively
speaking, it would be a Chan-Paton factor diagonal on the two endpoints).

\subsubsection*{Periodicities of forms}\label{s:cperiod}

The gauge transformations of the various potentials are
\begin{eqnarray}
\delta C_{(2)} &=& d\lambda_C \nonumber\\
\delta B_{(2)} &=& d\lambda_B \nonumber\\
\delta C_{(4)} &=& d\chi - \lambda_C \wedge H_{(3)}\ ,
\end{eqnarray}
in terms of one-forms $\lambda_C$ and $\lambda_B$ and three-form $\chi$.
The gauge transformation of $C_{(4)}$ corresponds to the field definition
$\tilde F_{(5)} = dC_{(4)} + C_{(2)} \wedge H_{(3)}$.  On $T^6$ these must
be periodic up to a gauge transformation,
\begin{eqnarray}
C_{(2)}(x+e^{m}) &=& C_{(2)}(x) + d\lambda^{m}_C(x) \nonumber\\
B_{(2)}(x+e^{m}) &=& B_{(2)}(x) + d\lambda^{m}_B(x) \nonumber\\
C_{(4)}(x+e^{m}) &=& C_{(4)}(x) + d\chi^{m}(x) - \lambda^{m}_C(x)
\wedge H_{(3)}(x)\ .
\end{eqnarray}
Here $e^{m}$ is the lattice vector in the $m$-direction, $(e^m)^n = 2\pi
\delta^{mn}$, and $\lambda^{m}_C$, $\lambda^{m}_B$,
and $\chi^{m}$ are specified gauge transformations.  To analyze these it
is convenient to write each field as its background value plus a shift,
for example $C_{(4)}(x) = \hat C_{(4)}(x) + c_{(4)}(x)$.
The three-form
flux backgrounds are constant, and so for the corresponding potentials we
can choose a gauge
\begin{equation}
\hat C_{mn} = \frac{1}{3} \hat F_{mnp} x^p\ ,\quad \hat B_{mn} = \frac{1}{3}
\hat H_{mnp} x^p\ .
\end{equation}
It follows that
\begin{equation}
\lambda^m_C = \frac{1}{6} \hat F_{mnp} x^n dx^p\ ,\quad \lambda^m_B =
\frac{1}{6} \hat H_{mnp} x^n dx^p\ .
\end{equation}
The quantized fluxes cannot fluctuate, and so the $\lambda^{m}_{B,C}$ are
fixed.  It then follows that the two-form fluctuations are periodic,
\begin{equation}
c_{(2)}(x+e^m) = c_{(2)}(x) \ ,\quad b_{(2)}(x+e^m) = b_{(2)}(x)\ .
\end{equation}

The four-form must satisfy a more complicated boundary condition.  This
can be deduced from the condition that $C_{(4)}(x + e^m + e^n)$ be
consistently defined, giving
\begin{equation}
d\chi^m(x+e^n) - d\chi^m(x) - d\chi^n(x+e^m) + d\chi^n(x) = \frac{1}{3}
\hat F_{nmq} dx^q \wedge H_{(3)}(x)\ .
\end{equation}
Note that it is the full $H_{(3)}$ that appears on the right-hand side, so
that $\chi^m$ has both a background piece and a field-dependent piece,
$\chi^m = \hat \chi^m + \zeta^m$.
Rather than solve directly for $\chi^m$ we first shift the four-form to
one with a simpler periodicity.  Define
\begin{equation}
\tilde c_{(4)} = c_{(4)} + \hat C_{(2)} \wedge b_{(2)}
+ \frac{1}{2} c_{(2)} \wedge b_{(2)}\ ,
\end{equation}
so that
\begin{equation}
\tilde f_{(5)} = d \tilde c_{(4)} - \Bigl(\hat F_{(3)}
+ \mbox{$\frac12$} f_{(3)}\Bigr) \wedge b_{(2)}
+ c_{(2)} \wedge \Bigl(\hat H_{(3)} + \mbox{$\frac12$} h_{(3)}\Bigr)\ .
\end{equation}
Then
\begin{equation}
\tilde c_{(4)}(x+e^m) = \tilde c_{(4)}(x) + d\tilde\zeta^m(x)\ ,
\end{equation}
where $\tilde\zeta^m = \zeta^m + \lambda^m_C \wedge b_{(2)}$.  It is
consistent to take $\tilde\zeta^m = 0$, and we choose a gauge in which
this is so.  A $\tilde\zeta^m$ that could not be gauged away would
correspond to a quantized five-form flux on $T^6$, which is inconsistent
with the ${\bf Z}_2$ projection~(\ref{reflect}).

The gauge variation of $\tilde c_{(4)}$ is
\begin{eqnarray}
\delta \tilde c_{(4)} &=& d\chi - \lambda_C \wedge H_{(3)} + \hat
C_{(2)} \wedge d \lambda_B + \frac{1}{2} (c_{(2)} \wedge d\lambda_B
+ d\lambda_C \wedge b_{(2)})
\nonumber\\
&=& d\tilde\chi - \lambda_C \wedge
\Bigl(\hat H_{(3)}  + \mbox{$\frac12$} h_{(3)}\Bigr) + \lambda_B \wedge
\Bigl(\hat F_{(3)}  + \mbox{$\frac12$} f_{(3)}\Bigr)
\nonumber\\ &=& d\tilde\chi - \frac{1}{2i\,{\rm Im}(\tau)}
\biggl\{ \OL{\lambda}_A \wedge \Bigl(\hat G_{(3)}  + \mbox{$\frac12$}
g_{(3)}\Bigr) - \lambda_A \wedge
\Bigl(\hat {\OL G}_{(3)}  + \mbox{$\frac12$} \OL g_{(3)}\Bigr) \biggr\}
 \ , \label{c4var}
\end{eqnarray}
where $\tilde\chi = \chi + C_{(2)} \wedge \lambda_B$
and $\lambda_A = \lambda_C - \tau \lambda_B$.
(Note that the
hatted background is defined to be fixed, so the gauge transformation goes
entirely into the fluctuation).  The gauge transformation $\tilde\chi$
must be periodic.  A nonperiodic gauge transformation would act on the
periodic identification by conjugation, $\tilde\zeta^{m\prime}(x) =
\tilde\zeta^m(x) + \tilde\chi(x+e^m) - \tilde\chi(x)$, so with fixed
identification the gauge transformation must be periodic.

\subsection{Gauge fields}\label{s:gauge}

The bulk vector fields that survive the orientifold
projection~(\ref{reflect}) are $c_{\mu n}$ and $b_{\mu n}$.  Form the
complex linear combinations
\begin{equation}
A_{\mu m} = C_{\mu n} - \tau B_{\mu n}\ .
\end{equation}
The gauge transformation is
$\delta A_{\mu m} = \partial_\mu \lambda_{A m}$,
where the one-form gauge parameter $\lambda_A$ is
as in eqs.~(\ref{c4varx}, \ref{c4var}).  It follows from the
transformation~(\ref{c4varx}) that the
$(1,0)$ parts of $\lambda_A$ leave the background invariant, so the unbroken
gauge fields are $A_{\mu i}$.  This is also evident from the linearized
gauge field strength
\begin{equation}
\tilde f_{(5)} = d \tilde c_{(4)} - \Bigl( a_{(2)} \wedge \hat {\OL
G}_{(3)} - \OL a_{(2)} \wedge \hat {
G}_{(3)} \Bigr)/2i\,{\rm Im}(\tau)\ . \label{elhiggs}
\end{equation}
The field $a_{\mu \bar\imath}$ appears in the $\mu \bar\imath jkl $
component.  Comparing with the nonlinear Higgs covariant derivative
$\partial_\mu \phi - A_\mu$, we see that $a_{\mu \bar\imath}$ is Higgsed by
$c_{\bar\imath jkl}$, so that $a_{\mu i}$ and $c_{\bar\imath \bar\jmath kl}$
remain as massless fields.

The real and imaginary parts of $a_{\mu i}$ give six
gauge fields; for example when $\tau = i$, these are
\begin{equation}
\frac{C_{\mu 4} - B_{\mu 7}}{\sqrt{2}}\ ,\ \
\frac{B_{\mu 4} + C_{\mu 7}}{\sqrt{2}}\ , \ \
\frac{C_{\mu 5} - B_{\mu 8}}{\sqrt{2}}\ , \ \
\frac{B_{\mu 5} + C_{\mu 8}}{\sqrt{2}}\ , \ \
\frac{C_{\mu 6} - B_{\mu 9}}{\sqrt{2}}\ , \ \
\frac{B_{\mu 6} + C_{\mu 9}}{\sqrt{2}}\ . \label{realvects}
\end{equation}
In addition each D3-brane adds a $U(1)$ gauge field, for total gauge group
$U(1)^{6 + N}$.  The total number of moduli is nine from the metric, nine
from $\tilde c_{(4)}$, and $6N$ from the D3-branes, for $6(3+N)$ in all.
The counting matches ${\cal N} = 3$ supergravity with $3+N$ matter
multiplets; note that this agreement requires exactly six of the $U(1)$'s
to be broken.

\subsubsection*{Massless vector solutions}

It is an interesting exercise, though somewhat aside from our main point,
to identify the massless vector solutions to the field equations, taking
into account the warping of the internal space.  We consider solutions
without D3-branes. We take as an ansatz that the only nontrivial components
of the fluctuations are the tensors,
$g_{\mu\nu m}$ and $\tilde f_{\mu\nu mnp}$.  The nontrivial field and
Bianchi equations are
\begin{eqnarray}
d g_{(3)} &=& 0 \ ,\quad
d * g_{(3)} = - i g_{\rm s} (g_{(3)} \wedge \hat{\tilde F}_{(5)}
+ \hat G_{(3)} \wedge \tilde f_{(5)})\ ,\nonumber\\
\tilde f_{(5)} &=& * \tilde f_{(5)} \ ,\quad d \tilde f_{5}
= \frac{i g_{\rm s}}{2} (\hat G_{(3)} \wedge \OL g_{(3)}
+ g_{(3)} \wedge \hat{\OL G}_{(3)})\ .
\end{eqnarray}
We further take
\begin{eqnarray}
g_{\mu\nu m} (x,y) &=& f_{\mu\nu}(x) u_m (y) + (*_4 f)_{\mu\nu}(x)
v_m(y)\ ,\nonumber\\
\tilde f_{\mu\nu mnp} &=& f_{\mu\nu}(x) \gamma_{mnp} (y) +
(*_4 f)_{\mu\nu}(x) (*_6 \gamma)_{mnp} (y)\ . \label{ansatz}
\end{eqnarray}
Here $u_m$ and $v_m$ are complex, and $\gamma_m$ and $f_{\mu\nu}$
are real.  In this subsection and the next we use $x$ for the noncompact
coordinates and
$y$ for the compact coordinates.  Subscripts of 4 (6) on the
Hodge star indicate that it is taken with respect to the spacetime
(internal) indices only.  Note that on two-forms $*_4$ is the same as the
flat spacetime Hodge star.

Inserting this ansatz into the field
equations gives the four-dimensional equations
\begin{equation}
d *_4 f_{(2)} = d f_{(2)} = 0
\end{equation}
and the internal equations
\begin{eqnarray}
d u_{(1)} &=& d v_{(1)} = 0 ,\nonumber\\
d *_6 u_{(1)} &=& - i g_{\rm s} v_{(1)} \wedge \hat{\tilde
F}_{(5)} - i g_{\rm s} \hat G_{(3)} \wedge *_6 \gamma_{(3)}\ ,
\nonumber\\
d *_6 v_{(1)} &=& i g_{\rm s} u_{(1)} \wedge \hat{\tilde
F}_{(5)} + i g_{\rm s} \hat G_{(3)} \wedge \gamma_{(3)}\ ,
\nonumber\\
d \gamma_{(3)} &=& \frac{i g_{\rm s}}{2} (\hat G_{(3)} \wedge
\OL u_{(1)} + u_{(1)} \wedge \hat{\OL G}_{(3)})
\ ,
\nonumber\\
d *_6 \gamma_{(3)} &=& \frac{i g_{\rm s}}{2} (\hat G_{(3)} \wedge
\OL v_{(1)} + v_{(1)} \wedge \hat{\OL G}_{(3)})
\ .
\end{eqnarray}
The Bianchi identities for $u_{(1)}$ and $v_{(1)}$ are solved by
\begin{equation}
u_{(1)}(y) = \omega_{(1)} + da(y)\ ,\quad v_{(1)}(y) =
\nu_{(1)} + db(y)
\end{equation}
where $\omega_{(1)}$ and $\nu_{(1)}$ are constant one-forms on the
internal space and
$a(y)$ and
$b(y)$ are periodic.
The equations for $\gamma_{(3)}$ are then solved by
\begin{equation}
\gamma_{(3)} = \frac{ig_{\rm s}}{2} (a \hat{\OL G}_{(3)} - \OL a
\hat G_{(3)})\ ,
\end{equation}
if
\begin{equation}
b(y) = - ia(y)\ ,\quad \omega_{\bar\imath} = \nu_{\bar\imath} = 0\ .
\end{equation}
Finally, the field equations for $u_{(1)}$ and $v_{(1)}$ both
become
\begin{equation}
Z \partial_m \partial_m a + 2 \partial_m Z \partial_m a + \partial_m Z
(\omega_m + i
\nu_m) = \frac{g_{\rm s}^2}{12} a \hat G_{mnp} \hat{\OL G}_{mnp}\ ,
\end{equation}
where all contractions are with the flat internal metric.  There are then
two solutions for each complex direction:
\begin{eqnarray}
\omega_{(1)} &=& - i \nu_{(1)} = dy^i\ ,\quad a = \gamma_{(3)} = 0\ ;
\quad
v_{(1)} = iu_{(1)}\ ,\nonumber\\
\omega_{(1)} &=& i \nu_{(1)} = dy^i\ ,\quad a\neq 0\ ,\quad\gamma_{(3)}
\neq 0 \ ;\quad
v_{(1)} = -iu_{(1)}\ .
\end{eqnarray}
For the second solution we do not have a closed form, but can show by a
variational argument that it exists.  Thus we have the expected six
internal solutions.
Note that we do not get distinct solutions by choosing $\omega_{(1)} = i
dy^i$, because the ansatz is invariant under $u \to v$, $v \to
-u$, $f_{(2)} \to *_4 f_{(2)}$.

\subsection{Metric on moduli space}\label{s:modmetric}

In this section, we will find the low-energy action for the scalars and
verify that it takes the form of a $U(3,n)/U(3) \times U(n)$ coset.  We
only consider the large-radius limit, where the warp factor $Z$ becomes
unity as discussed in section 2.3.
Thus we will drop the tildes
on the internal metric.  Four-dimensional geometric quantities will
be denoted by a ``4,'' or by ``E'' in the four-dimensional Einstein frame;
internal indices will always be raised with the string metric.

Let us first find the action for the metric moduli.  The dimensional
reduction of the ten-dimensional string frame Hilbert action gives
\be\label{metricmod1}
S_{g} = \frac{1}{4\pi \ap g_{\rm s}^2} \int d^4x \sqrt{-g_4}\,
\Delta \left[ R_4 +\Delta^{-2} \del_\mu \Delta \del^\mu \Delta
-\frac{1}{2} g^{\bj i}g^{\b{l} k} \del_\mu g_{k\bj} \del^\mu g_{i\b{l}}
\right]\ee
where $\Delta = \ap{}^{-3} \det g_{i\bj} $.
The dimensional reduction includes a factor $\frac{1}{2}(2\pi)^6$ from the
volume of $T^6/{\bf Z}_2$.  Switching to the
four-dimensional Einstein frame,
$2 g_{\rm s} g^{\rm E}_{\mu\nu}=\Delta g^4_{\mu\nu}$, the action becomes
\bea
S_{\rm g} &=& \frac{1}{2\pi \ap g_{\rm s}} \int d^4x \sqrt{-g_{\rm E}}
\left[ R_{\rm E} -\frac{1}{2}\Delta^{-2} \del_\mu \Delta
\del^\mu \Delta
-\frac{1}{2} g^{\bj i}g^{\b{l} k} \del_\mu g_{k\bj} \del^\mu g_{i\b{l}}
\right]\nonumber\\
&=& \frac{1}{2\pi \ap g_{\rm s}} \int d^4x \sqrt{-g_{\rm E}}
\left[ R_{\rm E} -\frac{1}{2} \gamma^{\bj i} \gamma^{\b{l} k} \del_\mu
\gamma_{k\bj}
\del^\mu \gamma_{i\b{l}} \right]\ , \label{metricmod2}\eea
where all spacetime indices are
raised with the Einstein metric.  We have
defined
\be\label{moddensity}
\gamma_{i\bj} = \frac{2 g_{\rm s}}{\alpha'} \frac{g_{i\bj}}{\Delta}
\ee
in order to eliminate double trace terms from the derivatives of $\Delta$;
the $\ap{}$ is included in order to make the moduli dimensionless.

The other bulk moduli are the R-R scalars, contained in the field strength
fluctuation $\tilde f_{(5)}$.  The moduli kinetic terms arise from $\tilde
f_{\mu npqr}$ and in Hodge dual form from $\tilde
f_{\mu \nu \lambda qr}$; in order to avoid the problems of self-dual
actions we include only the former, in terms of which
\begin{equation}
S_{\rm RR} = -\frac{ g_{\rm s}}{8\pi \alpha'}
\int d^4x \sqrt{-g_{\rm E}}\,| \tilde
f_{(5)} |^2 \ .
\end{equation}
In the absence of D3-branes, we have $f_{\mu ij \bar k \bar l} =
\partial_\mu c_{ij \bar k \bar l}$, and the action is simply
\begin{equation}
S_{\rm RR} = -\frac{g_{\rm s}}{32\pi \alpha'}
\int d^4x \sqrt{-g_{\rm E}}\, g^{i\bi'} g^{j\bj'} g^{k\bar k'} g^{l\bar l'}
\partial_\mu c_{ij\b{k}\b{l}}
\partial^\mu c_{\bi'\bj'k'l'}   \ .
\end{equation}
To exhibit the coset
structure we put these moduli in a two-index
form,
\be\label{defc_hat}
\ap c_{ij\= k \= l} = 2 \Delta^{-1} \epsilon_{ij\= k \= l a \= b}
\beta^{a \=b}\ .
\ee
The action for all the bulk supergravity moduli is then
\be\label{bulkmoduli}
S_{\mbox{\scriptsize mod}} =
- \frac{1}{4\pi \ap g_{\rm s}} \int d^4x
\sqrt{-g_{\rm E}}\, \gamma_{k\bar\jmath} \gamma_{i \bar l}
( \partial_\mu \gamma^{i\bj} \partial^\mu \gamma^{k\b{l}}
- \partial_\mu \beta^{i\bj} \partial^\mu \beta^{k\b{l}} )
\ .
\ee
This is just the $U(3,3)/U(3)\times U(3)$ moduli space metric, familiar
from the untwisted moduli of the ${\bf Z}_3$ orbifold 
\cite{Ferrara:1986qn,Ovrut:1989eq},
with upper and lower indices exchanged.

We now consider D3-branes.  Expanding the DBI action gives the kinetic term
\be\label{dbimod}
S_{\mbox{\scriptsize DBI}} = -\frac{1}{(2\pi)^3 \alpha' g_{\rm s}}
\int d^4x \sqrt{-g_{\rm E}}\,
\gamma_{i\bj} \del_\mu W_I^i \del^\mu \OL W_I^{\bj} \ ,
\ee
with an implicit sum on $I$.  In addition there is a dependence on the
collective coordinates from the coupling of the D3-brane to $C_{(4)}$, which
appears through a nontrivial five-form Bianchi identity.  In the D3-brane
rest frame,
\begin{equation}
d\tilde F_{(5)} = (2\pi)^4 \alpha'^2 \delta^6(y) d^6 y \to
\frac{\alpha'^2}{2\pi^2}  d^6 y\ ,
\end{equation}
where we have projected onto the zero mode; we omit the flux term in the
Bianchi identity, which makes no contribution to the moduli kinetic terms.
Boosting this gives
\bea
(d\tilde f)_{\mu\nu ij \bar k \bar l} &=& \frac{1}{2\pi^2 \alpha' \Delta}
\epsilon_{ij\bar k \bar l a \bar b} ( \del_{\mu}
W_I^{\bar b} \del_{\nu} W_I^a - \del_{\mu} W_I^a  \del_{\nu}
W_I^{\bar b} )\ , \nonumber\\
f_{\nu ij \bar k \bar l} &=& \partial_\nu c_{ij \bar k \bar l}
+ \frac{1}{4\pi^2 \alpha' \Delta}
\epsilon_{ij\bar k \bar l a \bar b} (
W_I^{\bar b} \del_{\nu} W_I^a - W_I^a  \del_{\nu} W_I^{\bar b} )\ .
\eea
The moduli space action is then
\be\label{allmoduli}
S_{\mbox{\scriptsize bulk}} =
- \frac{1}{4\pi \ap g_{\rm s}} \int d^4x
\sqrt{-g_{\rm E}}\, \biggl\{
\gamma_{k\bar\jmath} \gamma_{i \bar l}
( \partial_\mu \gamma^{i\bj} \partial^\mu \gamma^{k\b{l}}
- {\cal D}_\mu\beta^{i\bj} {\cal D}^{\mu}\beta^{ k\b{l}} )
+ \frac{1}{2\pi^2} \gamma_{i\bj}  \del_\mu W_I^i \del^\mu \OL
W_I^{\bj}
\biggr\}
\ ,
\ee
where
\be
{\cal D}_\mu\beta^{i\bj} = \partial_\mu \beta^{i\bj} + \frac{1}{8\pi^2} (
 W_I^{\bj} \del_{\mu} W_I^i - W_I^i \del_{\mu} W_I^{\bj} ) \ .
\ee

With a bit of algebra, it is possible to show that the entire action on
moduli space takes the form
\be\label{modaction}
S = \frac{1}{4} \frac{1}{2\pi\ap g_{\rm s}} \int d^4x \sqrt{-g_{\rm E}}
\,{\rm Tr} \left( \del_\mu M \eta \del^\mu M \eta\right)
\ee
where $\eta$ is the $U(3,3+N)$ invariant metric ($\eta =\Omega^\dagger
\eta \Omega$) and $M$ is a Hermitean
$U(3,3+N)$ matrix that behaves as $M\rightarrow \Omega M \Omega^\dagger$
under $U(3,3+N)$.  We work in a basis with block diagonal form
\be\label{matrices}
\eta = \left[ \begin{array}{ccc} &\, I_3\, &\\ \, I_3\, &&\\&&\, I_N\, \ea
\right]\ ,\quad M = \left[ \begin{array}{ccc}\gamma^{-1} &
-\gamma^{-1}\mathcal{B}&  -\gamma^{-1} \alpha^\dagger\\
-\mathcal{B}^\dagger \gamma^{-1} &\ \gamma +\mathcal{B}^\dagger\gamma^{-1}
\mathcal{B} +\alpha^\dagger \alpha\
& \mathcal{B}^\dagger\gamma^{-1} \alpha^\dagger +\alpha^\dagger\\
-\alpha \gamma^{-1} & \alpha\gamma^{-1} \mathcal{B}+\alpha&
I_N+\alpha\gamma^{-1} \alpha^\dagger\ea
\right]
\ee
with matrix notation $\gamma = \gamma^{\bj i}$,
$\alpha = W^i_I/2\pi$, and
$\mathcal{B}=\beta + (1/2)\alpha^\dagger \alpha$.  To verify that this takes
the appropriate coset form, note that we can write
\be\label{coset}
M=V^\dagger V\ ,\quad V = \left[\begin{array}{ccc} e& -e\mathcal{B}&
-e\alpha^\dagger\\
0&e^{-1}& 0\\0&\alpha&I_N\ea\right]
\ee
where $e$ is the vielbein $e^\dagger e = \gamma^{-1}$.  Following
\cite{Maharana:1993my}, we see that $M$ indeed belongs to the coset
$U(3,3+N)/U(3)\times U(3+N)$, precisely as we expected based on $\N=3$
supersymmetry.

\subsection{Comparison to $\N=4$ heterotic string}\label{s:heterotic}

The results of \S\ref{s:modmetric} are notably similar to work done
by Maharana and Schwarz on the $O(6,22)$ duality of the heterotic string
on $T^6$ \cite{Maharana:1993my}.  This is not an accident.  Starting from
the heterotic string, $S$-duality maps to type I strings, and a further
$T$-duality on all six dimensions takes the theory to the IIB model of
\cite{Verlinde:1999fy}.  Our $\N=3$ models are then obtained by
nonperturbatively transforming D3-branes into self-dual $G_{(3)}$ flux,
so we expect that our moduli space should simply be a subspace of the
heterotic moduli space.

To make this more precise, we can follow the action of the $S$- and
$T$-dualities on the moduli of the heterotic theory.  For ease of 
comparison, we will use coordinates of radii equal to the string length
$\sqrt{\ap}$.  We will also choose duality conventions such that $\ap$ 
is the same in the heterotic, type I, and type IIB string theories.  To
get the normalization correct including numerical factors, we must be
careful (see \cite{Gimon:1996rq} for some factors in the type I theory,
for example).

We start by considering the heterotic--type I S-duality.  Under this
duality, the heterotic fundamental string maps to the type I D-string; 
in particular the actions must be equal.  Since the D-string tension
and charge are reduced by a factor of $\sqrt{2}$ by the orientifold 
projection in the type I theory, we therefore must have
\bea
\frac{1}{2\pi\ap \sqrt{2}} \int d^2\xi e^{-\Phi}({\rm I}) 
\sqrt{-\det g({\rm I})} &=& \frac{1}{2\pi\ap} \int d^2\xi
\sqrt{-\det g({\rm het})}\nonumber\\
\Rightarrow g_{MN}({\rm het}) &=&
\frac{e^{-\Phi}({\rm I})}{\sqrt{2}} g_{MN}({\rm I}) \label{hetI}
\eea
and likewise $B_2({\rm het}) = C_2({\rm I})/\sqrt{2}$.
The 10D supergravity actions then map into each other if we take the 
gauge theory potentials to be equal.

In the T-duality between type I on $T^6$ and IIB on 
$T^6/\mathbf{Z}_2$, the dilaton picks up a well-known factor of
$\sqrt{2}$ \cite{Gimon:1996rq}, so the T-duality is
\be\label{IIIBdilg}
e^{\Phi}({\rm I}) = \frac{\sqrt{2}}{\det^{1/2} g_{mn}}
e^{\Phi}({\rm IIB})\ , \quad
g_{mn}({\rm I}) = g^{mn}({\rm IIB}) \ ,\quad
g_{\mu\nu}({\rm I}) = g_{\mu\nu}({\rm IIB})\ .\ee
There is an additional factor in the RR sector, as follows.  Taking the
prefactor of the 10D action to be the same in the two theories, 
T-duality tells us that we should have the same dimensionally reduced
actions, or
\be\label{IIIBRRaction}
\frac{(2\pi)^6\ap{}^3}{2\cdot 2}\int d^4x\sqrt{-g_4}\Delta 
\del_\mu C_{mn} \del^\mu C^{mn} ({\rm I})
= \frac{(2\pi)^6\ap{}^3}{2\cdot 2\cdot 4!}\int d^4x\sqrt{-g_4}
\Delta \del_\mu C_{mnpq} \del^\mu C^{mnpq} ({\rm IIB})\ee
for the moduli.  Here, $\Delta=\det^{1/2}g_{mn}$ and $g_4$ is the 
string frame metric.  The additional factor of 2 in the IIB case 
again comes from the volume.  This equality holds if we take
\be\label{IIIBRR}
C_{mn}({\rm I}) = \frac{1}{\sqrt{2}\cdot 4!} \Delta \epsilon^{mnpqrs}
C_{pqrs}\ .\ee

Then the heterotic moduli 
(using the notation of \cite{Maharana:1993my}) map
to the IIB $\N=4$ moduli as follows:
\be
\label{STmap}
g_{\mu\nu} \rightarrow g_{{\rm E}\, \mu\nu}\ ,\quad g_{mn}\rightarrow
\gamma^{-1}{}^{mn}\ ,\quad B_{mn}\rightarrow \beta^{mn}\ ,
\quad a_m^I\rightarrow \alpha^m_I\ ,
\ee
following the notation of \S \ref{s:modmetric} for the IIB side, up to
factors of $\ap$ from coordinate rescaling.
The $\N=3$ moduli are then clearly the (anti-)Hermitean subset of the
gravitational and R-R moduli along with all the D-brane positions in
complex form.

There is an additional complex modulus
in the $\N=4$ case which corresponds on the heterotic side to the
four-dimensional dilaton and $B_{\mu\nu}$ axion, and on the IIB side to the
ten-dimensional dilaton and R-R scalar.  In the $\N=3$ theories this
modulus is fixed.

Consider the ${\cal N}=4$ states which become massive due to the
fluxes.  These include one gravitino, so we must have a massive spin-3/2
multiplet.  This must be a large representation because these supergravity
states are all neutral under the $U(1)$ central charges, and so the
helicities are
\be\label{spin32}
\textstyle\frac{3}{2}\
,\ 1^6\ ,\ \frac{1}{2}^{15}\ ,\ 0^{20}\ ,\ -\frac{1}{2}^{15}\ ,\ -1^6\ ,\
-\frac{3}{2}\ .
\ee
This agrees with the finding that six gauge symmetries are broken.  The
twenty spin-zero components are the dilaton-axion, the six zero-helicity
components of the massive vectors (from $C_{(4)}$), and the twelve
real components of
$g_{w^i w^j}$.  Note that at large radius these states, with masses
$\alpha'^2/R^6$, lie parametrically below the Kaluza-Klein scale of
$R^{-2}$.  Thus we can truncate to an effective field theory in which only
these and the massless states survive.  Since the mass scale is
parametrically below the Planck scale as well, the  SUSY breaking from
$\N=4$ to $\N=3$ must be spontaneous.  There has been some discussion of
such breaking in
supergravity~\cite{deRoo:1986yw,Wagemans:1988zy,Tsokur:1996gr}.

\sect{Dualities}\label{s:dualities}

In this section, we discuss the stringy duality group of these
compactifications.  In particular, we are interested in the dual
description that governs the physics when the radii become small.

\subsection{Dualities of the $\N=4$ theory with 16 D3-branes}

As a warmup, let us first consider the dualities of the $\N=4$ theory with
16 D3-branes, which is the $T$-dual of type I on $T^6$ and the $TS$-dual
of the heterotic theory on $T^6$.  The duality of the latter theory is
$SO(22,6,{\bf Z}) \times SU(1,1,{\bf Z})$~\cite{Schwarz:1993mg}.
Consider first the  perturbative $SO(22,6,{\bf Z})$ factor.  This group is
generated by discrete shifts of the Wilson lines, Weyl reflections in the
gauge group, discrete shifts of
$B_{mn}$, large coordinate transformations on the torus, and the inversion
of one or more directions on the torus (this is not meant to be a minimal
set of generators).  We will call this last operation
$R$-duality to distinguish it from the full perturbative $T$-duality.  The
first three operations are manifest in the IIB description, as the
periodicities of the D3-brane collective coordinates, permutations
of the D3-branes, discrete shifts of the $C_{mnpq}$, and large coordinate
transformations respectively.  The
$R$-duality is not manifest in the IIB description.  Note that this is not
the same as IIB
$R$-duality, because it leaves fixed the ten-dimensional IIB coupling and
not the four-dimensional coupling.  Rather, it is the image of the
heterotic $R$-duality; therefore we will henceforth designate it $R_{\rm
het}$.

To see $R_{\rm het}$ in the IIB description it is useful to focus on its
action on the BPS states.  In the heterotic description
$R_{\rm het}$ interchanges KK states and winding F-strings.  In the type I
description these become KK states and winding D-strings, and then in type
IIB they become winding F-strings and D5-branes.  Similarly it interchanges
winding D-strings and NS5-branes.

To analyze the duality carefully we need the masses of these
objects, taking for simplicity a rectangular torus $ds^2 = r_m^2 dx^m
dx^m$, and vanishing R-R backgrounds. We take the F- and D-strings to be
wound in the 4-direction, and the D5- and NS5-branes to be wound in the
56789-directions.  Then (in the string frame)
\bea
m_{\rm F1} &=& \frac{r_4}{\alpha'} \ ,\quad
m_{\rm D1} = \frac{r_4}{\alpha' g_{\rm s}}\ ,
\nonumber\\
m_{\rm D5} &=& \frac{v}{2 r_4 \alpha'^3 g_{\rm s}}
\ ,\quad
m_{\rm NS5} = \frac{v}{2 r_4 \alpha'^3 g^2_{\rm s}}
\ ,
\eea
where $v = \prod_m r_m$.  The factors of 2 come about because the
strings must be wound on cycles of $T^6$, while the 5-branes can be wound
on the fixed cycle $x^4 = 0$ whose volume is halved.  For the F-string
this represents the fact that in an orientifold the closed strings are
obtained by projection; for the NS5-brane it is simply the ${\bf Z}_2$
reduction of an NS5-brane solution at $x^4 = 0$ on the original $T^6$.
For the D1- and D5-branes, these statements are $T$-dual to the fact that
in the type I string the D5-brane has two Chan-Paton values while the
D1-brane has one~\cite{Witten:1996gx,Gimon:1996rq}: thus, the IIB D1-brane
can move off the fixed plane, while the D5-brane is fixed.
For future reference let us also give the masses in the type I description,
where $r_m' = \alpha'/r_m$; the couplings are related by $v' / g'^2_{\rm s}
= v/2g_{\rm s}^2$, the factor of 2 being from the orientifold volume.  Then
\be
m_{\rm KK'} = \frac{1}{r_4'} \ ,\quad
m_{\rm D5'} = \frac{v' \sqrt{2}}{r_4' \alpha'^3 g'_{\rm s}}\ ,
\quad
m_{\rm D1'} = \frac{r'_4}{v' g'_{\rm s} \sqrt{2}}
\ .
\ee
The factors of $\sqrt 2$ are as found in ref.~\cite{Gimon:1996rq}.

In units of the four-dimensional Planck mass $m_4 = (v/2)^{1/2}
\alpha'^{-2}g_{\rm s}^{-1}$ the BPS masses are
\bea
\frac{m_{\rm F1}}{m_4} &=& \frac{r_4 \alpha' g_{\rm s}
\sqrt{2}}{v^{1/2}}
= \frac{g^{1/2}_{\rm s}}{\rho_4}\ ,\quad
\frac{m_{\rm D1}}{m_4} = \frac{r_4 \alpha' \sqrt{2}}{v^{1/2}}
= \frac{1}{\rho_4 g_{\rm s}^{1/2}}\ , \nonumber\\
\frac{m_{\rm D5}}{m_4} &=& \frac{v^{1/2}}{r_4 \alpha' \sqrt{2}}
= \rho_4 g^{1/2}_{\rm s}\ ,\quad
\frac{m_{\rm NS5}}{m_4} = \frac{v^{1/2}}{r_4 \alpha' g_{\rm
s}\sqrt{2}}
= \frac{\rho_4 }{g_{\rm s}^{1/2}}\ . \label{n4mass}
\eea
We have defined $\rho_4 = {v^{1/2}}/r_4 \alpha' g^{1/2}_{\rm s}
\sqrt{2}$, which is just the radius in the heterotic string picture, in
heterotic string units.  The first and second lines interchange under
inversion of $\rho_4$, as expected.

The $SU(1,1,{\bf Z})$ of the heterotic theory maps to the $SU(1,1,{\bf Z})$
of the ten-dimensional IIB theory.  In particular, $g_{\rm s} \to g_{\rm
s}^{-1}$ interchanges the states in each line of eq.~(\ref{n4mass}).

\subsection{Dualities of the $\N = 3$ theories}\label{s:bps}

We expect that the duality group will be an integer version of the
continuous low energy symmetry $U(3,3+N)$.  The simplest guess would be
that it is the intersection of this continuous group with the discrete
symmetry $SO(6,22,{\bf Z}) \times SU(1,1,{\bf Z})$ of the $\N = 4$ theory.
In other words, the fluxes break the duality symmetry to a subgroup, just as
they do with the supersymmetry.  However, we will see that this guess is
incorrect.

Let us consider the BPS states discussed in section~4.1.
Note that these do not have a perturbative description, because $g_{\rm
s}$ is of order one, but we can study them using the effective low energy
description when the radii are large.  In the $\N = 4$ theory, these
states are invariant under eight supersymmetries; one finds that four of
these supersymmetries lie in the $\N=3$ subalgebra of interest.\footnote
{More details, and further analysis, will be presented in future work.}
Thus  these are ``$1/3$-BPS'' states, in agreement with the result that BPS
particles in $\N=3$ preserve four supersymmetries
\cite{Kounnas:1998hi}.

When the torus is rectangular, the R-R backgrounds
zero, and all D3-brane coincident, the central charges are from the bulk
$U(1)$'s $A_{\mu i}$.  For simplicity let us focus on the case that
$g_{\rm s} = 1$.  The unbroken gauge fields associated with the 4-7
torus are
\be
\frac{B_{\mu 4} + C_{\mu 7}}{\sqrt{2}}\ ,\quad
\frac{C_{\mu 4} - B_{\mu 7}}{\sqrt{2}}\ , \label{unbroken}
\ee
while the broken symmetries are
\be
\frac{B_{\mu 4} - C_{\mu 7}}{\sqrt{2}}\ ,\quad
\frac{C_{\mu 4} + B_{\mu 7}}{\sqrt{2}}\ .
\ee
Thus a D-string in the 4-direction, or an F-string in the 7-direction,
have the same BPS charge, electric charge in the first $U(1)$.  A D5-brane
in the 56789-directions, and an NS5-brane in the 89456-directions, carry
the analogous magnetic charge.\footnote{More generally we can consider
$(p,q)$-strings and 5-branes, at various angles --- a full accounting of
the BPS states is an interesting exercise.}

There is, however, an important subtlety: not all of these states
actually appear in the spectrum.  Each of these objects couples both to a
massless and a massive vector.  The discussion of eq.~(\ref{elhiggs})
shows that the vector mass arises from electric Higgsing.
For the electrically charged 1-branes the massive charge is screened and
there is no great effect.  However, the 5-branes carry the corresponding
magnetic charge and so must be confined: the Higgsing breaks the symmetry
between these two sets of states.  We can understand this in two other
ways as well.  First, the Higgsing reduces the long-ranged interaction
between the electric and magnetic objects by a factor of two.  Since they
had the minimum relative Dirac quantum in the $\N = 4$ theory, the are no
longer correctly quantized.  Second, the gauge invariant flux on the
D5-brane is
$\mathcal{F}_{(2)}=F_{(2)}-B_{(2)}/2\pi\alpha'$, which satisfies
\be
d\mathcal{F}_{(2)} = -H_{(3)}/2\pi\alpha'\ . \label{dfish}
\ee
The integral of this over any
3-cycle should then vanish, but this is inconsistent because our background
includes at least one of $H_{678}$ or $H_{567}$, among others.
In order that the Bianchi identity be consistent, there must be another
source.  This would be a D3-brane, which is localized in the 3-cycle in
question and extended in the other two compact directions and one
noncompact direction: this is a confining flux tube.

It follows that the duality $R_{\rm het}$ that interchanges the basic 1-
and 5-branes does not survive in the $\N=3$ theory.\footnote
{Note that this duality interchanges electric and magnetic
objects, while the $SO(6,22,{\bf Z})$ of the heterotic theory acts
separately on each. This is because the unbroken gauge
fields~(\ref{unbroken}) are a linear combination of electric and magnetic
gauge fields in the heterotic picture: the nonlinear Higgs field has both
electric and magnetic charges. \label{magfoot} }
There are
magnetic objects in this theory, but they are bound states.  For example, a
56789 D5-brane and a 89456 NS5-brane have the same BPS $U(1)$ charge and
the opposite broken charge, and so their bound state is unconfined and is
a BPS state of twice the minimum $\N = 4$ mass.  In a perturbative
description, the D5-brane ends twice on the NS5-brane, as in the
$(p,q)$-5-brane webs of \cite{Aharony:1997ju,Aharony:1998bh}.

The simplest conjecture would then be that the duality group interchanges
the objects of minimum electric and magnetic charge.  With the D5-brane
masses~(\ref{n4mass}) doubled, this would now mean that $\rho'_m =
1/2\rho_m$; it is not clear whether this symmetry could be inherited from
the $\N=4$ theory.\footnote{Such a duality does exist in the heterotic
string for a nonzero axion \cite{Witten:1998bs}, but it has not been
determined if
it can be combined with the heterotic strong-weak coupling duality 
\cite{Sen:1994fa} to generate the proper action on the BPS states. 
This possibility also requires that the axion of the ``heterotic''
description of the $\N=3$ theory be shifted by half a unit, and it
is not immediately clear that this is so.}
To be precise, this symmetry can act independently on any set of paired
indices, 4-7, 5-8, or 6-9: it must preserve eq.~(\ref{pairrad}).  This
conjectured symmetry relates rather different objects, and so for example
the total number of BPS states of a D-string in the 4-direction and an
F-string in the 7-direction must equal that of the D5/NS5 bound states.
It is an interesting exercise, to be studied in
future work, to determine the BPS spectra of these objects as a function
of the background fluxes.  It is possible that this will reveal a more
intricate pattern of dualities, in which the various $\N=3$ models mix.
It is conceivable that the dualities might involve other types of $\N=3$
construction, such as those of ref.~\cite{Ferrara:1989nm}, though we have
no particular reason to expect this.  Note also that there is no reason to
expect an effective heterotic description anywhere in the moduli space.
For the $\N=4$ theories such a description holds when the IIB radii are
small and the ten-dimensional IIB coupling is large, but in the $\N=3$
models the latter coupling is always of order one.

The remainder of the duality group would be generated by large coordinate
transformations mixing the holomorphic coordinates, periodicities of the
D3-brane coordinates, permutations of the D3-branes, and shifts of the R-R
backgrounds.  We conclude this section with a few remarks about these.

When discussing large coordinate transformations on the torus, we should
distinguish between U-dualities, which leave the background invariant, 
and ``string-string''-like dualities, which take one background into
a different but equivalent background.  The transformations 
that give ``string-string'' dualities are discussed
in \cite{Kachru:2002he}; here we are interested in finding those that
give U-dualities.

A large coordinate transformation will leave the background $G_{\b 1\b 2\b
3}$ invariant if its determinant is unity.  Nevertheless, the duality
also includes elements of nontrivial determinant.  For example, at $\tau=i$,
rotation of a single coordinate
$w^1\to iw^1$ changes the background 3-form $G_{\b 1\b 2\b 3}\to iG_{\b
1\b 2\b 3}$, but this can be undone by one of the broken
$SL(2;\mathbf{Z})$ dualities of the IIB string, $\tau\to -1/\tau$.
Note that this combined operation leaves the background
invariant and so does not act on the moduli space, but it does mix the
BPS states and so is a nontrivial duality.  Also, if the fluxes are
chosen so that $\tau\neq i$, this duality is not a U-duality, so we
find that different $\N=3$ backgrounds have slightly different U-duality 
groups.
Note that in models with fluxes on the orientifold planes,
we must restrict to transformations that take
O3-planes of a given type into the same type.  If we insist
that all the fixed points map to themselves under dualities, then the
off-diagonal elements of the linear transformation must be even and the
diagonal elements must be unity (or $-1$ with a translation).  Again,
different backgrounds will have different U-duality groups.

The D3-brane gauge charges do not appear in the IIB superalgebra, and a
zero-length F- or D-string stretched between coincident D3-branes is
massless, giving an enhanced gauge symmetry.  When the D3-branes are
separated the stretched string begins to couple to the bulk gauge fields,
and acquires a BPS mass and charge.  When the D3-branes shift fully around
the 1-cycles of the torus, the attached F- and D-strings acquire integer
winding charges.  Since the electric charges on the D3-branes are the end
points of F-strings, this duality shifts the
bulk electric charges by the D3-brane electric charges.   Note that
since the magnetic D3-brane charges are D-string end points, the shift
also depends on the D3-brane magnetic charges: as noted in
footnote~\ref{magfoot}, the duality group is nontrivially embedded in the
low energy electric/magnetic duality group.

In order to understand the R-R shift dualities in detail one needs to
consider two other classes of BPS objects.  The first are Euclidean
D3-branes wrapped entirely on the internal torus.  These are instantons
under the unbroken gauge symmetries, and their phases depend on the R-R
moduli.  The magnetic analogs to these are spacetime strings,
D3-branes wrapped on the appropriate 2-cycles of the torus and
extended in one direction of the external space; we have already
encountered these above, as confining flux tubes.  As one circles such a
string one traces a closed loop in moduli space.  The discrete shift
dualities must leave all instanton amplitudes invariant, and one expects
that all such shifts will be generated by the dual strings.  Note that
the instantons wrap enough directions for the identity~(\ref{dfish}) to be
relevant, so their spectrum will be subject to restrictions.

There are two physically distinct cases of these instantons and strings.
The simpler case couple to the diagonal $\beta^{\bi i}\ (i=\bi)$ moduli,
as these moduli correspond to a single real component of $\tilde c_{(4)}$.
For example, $\beta^{\b 1 1}$ couples to an instantonic D3-brane wrapped
on the $5689$ directions and a string D3-brane partially wrapped on
the $47$ directions.  Notice that these instantons do not wrap any 3-cycle
including $H_{(3)}$ or $F_{(3)}$ flux.  Additionally, we have checked that
these strings preserve supersymmetry; in fact, they preserve 6 supercharges
in common with the background.  The other case correspond to the off-diagonal
moduli, which has real and imaginary parts constructed from two
components of $\tilde c_{(4)}$ each.  The instantons do wrap 3-cycles
with flux, so they must come in bound states, much as the magnetic BPS
charges discussed above, and the corresponding strings would then
fill half a supermultiplet each.  These strings preserve four
supersymmetries in common with the background.

We consider here just the diagonal case.  In the $\N=4$ theory the wrapped
D3-branes are dual to type I instantonic D-strings.   These
have a single Chan-Paton index, so there exist D3-brane instantons
wrapping one of the special half-volume 4-cycles.  Their action is given by
\be\label{c4inst}
\frac{1}{(2\pi)^3\ap{}^2} \int \tilde c_{5689} dx^5dx^6dx^8dx^9
= \frac{\pi}{\ap{}^2} \tilde c_{5689}\ .\ee
This implies that $\tilde c_{5689}$ can shift by even integral multiples
of $\ap{}^2$ without changing the path integral.  As this shifts
$\beta^{\b 1 1}$ by $i/2$ times that integer, we see that the shift
duality has been broken by the instantons to $\mathbf{Z}$ for each
axion.

Let us check that this is consistent with the spacetime strings.  A
D3-brane wrapped on $47$ is dual to a D5-brane in the
$\N=4$ type I theory.  Since the type I D5-brane must have
two Chan-Paton indices, these D3-branes can only wrap 2-cycles of volume
$(2\pi)^2$.  Using the relative coefficients of terms in the action,
the 10-dimensional Bianchi identity for the 5-form integrates to
\be\label{5bianchi}
\frac{1}{(2\pi)^7\ap{}^4} \oint_M \tilde F_{(5)} =
\frac{1}{(2\pi)^3\ap{}^2}\ .
\ee
The surface surrounding the string is $M=S^1\times T^4/
\mathbf{Z}_2$.  Integrating over the latter factor gives
\be
\oint_{S^1} d\tilde
c_{5689} = 2\ap{}^2\ ,
\ee
which is the minimum shift consistent with the instanton amplitude.

A complete analysis of the duality group is left for future work.

In conclusion, we see that although supersymmetry strongly constrains these
$\N=3$ models, there remain interesting dynamical issues.  Thus these
models may be a useful preliminary to the study of less symmetric and more
realistic warped compactifications.

\subsection*{Acknowledgements}

We would like to thank A. Hanany, S. Sethi, and E. Witten for helpful
discussions and communications.  This work was supported by NSF grants
PHY99-07949 and PHY00-98395.
The work of A. F. was supported in part by a National Science Foundation
Graduate Research Fellowship.

\bibliographystyle{h-physrev4}
\bibliography{compact3e}

\end{document}